**Hatchery-produced sandfish (*Holothuria scabra*) show altered genetic diversity in New Caledonia**

Florentine RIQUET[1,2], Cécile FAUVELOT[1,2], Pauline FEY[1], Daphné GRULOIS[1], Marc LEOPOLD[1,3]

**Affiliations and addresses:**

[1] UMR ENTROPIE (IRD, University of La Reunion, CNRS, University of New Caledonia, Ifremer), 98800 Noumea, New Caledonia

[2] UMR LOV (Sorbonne University, CNRS), 06230 Villefranche-sur-Mer, France

[3] UMR ENTROPIE (IRD, University of La Reunion, CNRS, University of New Caledonia, Ifremer), 97400 Saint-Denis, La Reunion c/o IH.SM, University of Toliara, BP 141, 601 Toliara, Madagascar

**E-mail adresses:**

Florentine Riquet: florentine.riquet@ird.fr; Cécile Fauvelot: cecile.fauvelot@ird.fr; Pauline Fey: pauline.fey@gmail.com; Daphné Grulois: daphne.grulois@hotmail.com; Marc Léopold: marc.leopold@ird.fr

**Corresponding author:** Florentine Riquet, flo.riquet@gmail.com , +33 (0)659142410, Laboratoire d'Océanographie de Villefranche (LOV) - UMR 7093 UMPC/CNRS, 181 chemin du Lazaret, 06230 Villefranche-sur-Mer, France

**Abstract**

Facing an alarming continuing decline of wild sea cucumber resources, management strategies were developed over the past three decades to sustainably promote development, maintenance, or regeneration of wild sea cucumber fisheries. In New Caledonia (South Pacific), dedicated management efforts via restocking and sea ranching programs were implemented to cope with the overharvesting of the sandfish *Holothuria scabra* and the recent loss of known populations. In order to investigate genetic implications of a major *H. scabra* restocking program, we assessed the genetic diversity and structure of wild stocks and hatchery-produced sandfish and compared the genetic outcomes of consecutive spawning and juvenile production events. For this, 1358 sandfish collected at four sites along the northwestern coasts of New Caledonia, as well as during five different restocking events in the Tiabet Bay, were genotyped using nine polymorphic microsatellite markers. We found that wild *H. scabra* populations from the northwestern coast of New Caledonia likely belonged to one panmictic population with high level of gene flow observed along the study scale. Further, this panmictic population displayed an effective size of breeders large enough to ensure the feasibility of appropriate breeding programs for restocking. In contrast, hatchery-produced samples did suffer from an important reduction in the effective population size: the effective population size were so small that genetic drift was detectable over one generation, with the presence of inbred individuals, as well as more related dyads than in wild populations. All these results suggest that dedicated efforts in hatcheries are further needed to maintain genetic diversity of hatchery-produced individuals in order to unbalance any negative impact during this artificial selection.

## 1. Introduction

To meet the rising demand for trepang or *bêche-de-mer* in the Asian market, sea cucumber fisheries expanded in the 1990s and progressively spread to most tropical countries, while exacerbating fishing pressure and generating major resource sustainability issues (Anderson et al., 2011; Purcell et al., 2013). Wild tropical sea cucumbers are particularly vulnerable to overfishing, as they are easy to harvest in shallow waters. Further, when heavily exploited, wild populations usually show low recovery rate, with scattered spawners impairing the species reproductive success (Uthicke and Benzie, 2001; Uthicke and Conand, 2005). As a result, most sea cucumber stocks keep declining worldwide and show decreasing recovery rate despite implemented management strategies to sustain fisheries, as for instance the fishing moratoria adopted in the Indo-Pacific region (Baker-Médard and Ohl, 2019; Conand, 2018; Léopold et al., 2015; Purcell et al., 2013; Uthicke and Conand, 2005). Although ecological function of sea cucumber species in coral reef ecosystems and associated soft substrates are well established (Purcell et al., 2016), knowledge gaps in the biology and the ecology (e.g. growth rate, mortality, and juvenile recruitment) of most sea cucumbers trigger wild population dynamics difficult to predict and, therefore, to manage (Conand, 2018).

Facing the rapid depletion of wild sea cucumber resources, technical advances in artificial reproduction and farming of few high-value species of holothurians have considerably increased over the past three decades (Lovatelli, 2004). Releasing cultured benthic invertebrates such as sea cucumbers in the wild was viewed as a possible strategy to promote development, sustainability, or regeneration of fisheries at national or global scale (Bell et al., 2005). Specifically, the sandfish *Holothuria scabra* has attracted most research effort in the Indo-Pacific region (Bermudes, 2018; Hair et al., 2012, 2016; Lovatelli, 2004) due to its high commercial value (e.g. Purcell et al., 2018) and its high potential for restocking (Battaglene, 1999; Bell et al., 2005). Nevertheless, aquaculture-based stock enhancing strategies may alter genetic diversity of wild populations, depending on the number and geographical origin(s) of the breeders and the highly unpredictable number of released juveniles that will survive and reproduce (Hair et al., 2016; Purcell et al., 2012; Utter, 1998). Releasing genetically similar sea cucumber juveniles has been recommended to reduce both the introduction of exogenous alleles and outbreeding depression in the target wild population (Cross, 2000; Uthicke and Purcell, 2004; Ward, 2006). In that perspective, several studies have recently assessed the genetic diversity and structure of wild sandfish resources to inform restocking projects (Ninwichian and Klinbunga, 2020; Nowland et al., 2017; Ravago-Gotanco and Kim, 2019). However, to our knowledge, genetic diversity of wild stocks, broodstock, and hatchery-produced juveniles of sandfish has not been concurrently investigated, although such information is critical to assess the genetic effects of hatchery and restocking initiatives, especially in this species listed in the IUCN Red List of Threatened Species (Hamel et al., 2013).

To address that concern, we investigated the genetic implications of distinct restocking events of *H. scabra* in New Caledonia (South Pacific). Exploitation of sea cucumbers in New Caledonia initiated in the 1840s following trade opportunities in China (Conand, 1990). Increasing harvest severely affected sandfish resources in the 2000s (Purcell et al., 2012), which prompted the provincial fisheries divisions to implement management regulations (e.g. Léopold et al., 2013) while seeking for sandfish stock enhancement opportunity in shallow, inshore and depleted areas (Purcell et al., 2002). Genetic homogeneity of sandfish populations was observed along the northwestern coast of New Caledonia using allozyme polymorphism (Uthicke and Purcell, 2004), suggesting that releasing hatchery-produced juveniles from broodstock collected within that area would not cause adverse genetic effects. These results encouraged the fisheries division of the Northern Province of New Caledonia to plan reseeding of hatchery-produced juveniles in the Tiabet area (Northern Province, New Caledonia), where sandfish had locally disappeared due to overfishing. From 2011 to 2014, more than 100 000 hatchery-produced juveniles were released in large sea pens (from 2011 to 2013) or open habitat (in 2014) for restocking the Tiabet Bay. In the present study, we 1) examined the genetic diversity and structure of sandfish wild individuals, including broodstock, and hatchery-produced individuals using highly polymorphic microsatellite markers, and 2) compared the genetic outcomes of different restocking events. Recommendations for future sandfish restocking, stock enhancement and fishery management are eventually discussed based on the survey results.

## 2. Materials and Methods

### 2.1 Sampling design

The survey was conducted in the northwestern lagoon area of New Caledonia, from Kone to the north tip of Grande Terre, the Tiabet Bay, chosen as the restocking area consequently to the collapse of wild *H. scabra* population (Fig. 1). Wild and cultured sea cucumbers were all sampled in 2014 through tegument biopsies and stored in 80% EtOH for genetic study as follows.

First, to assess the spatial genetic diversity and structure of the wild sandfish population at the study scale (ca. 90 km), 400 adults were collected at four sites along the northwestern lagoon of New Caledonia (100 individuals per site), where suitable habitats for sandfish are found on coastal reef flats and seagrass beds. Adults were sampled in the lagoons of Koumac, Tinip, Boyen and Kone (from north to south, sites 1 – 4, respectively, Fig. 1) where local fishermen regularly harvest wild populations of *H. scabra*.

Second, to assess genetic effects of restocking events, samples from five distinct batches of hatchery-produced individuals were collected as follow:

- Three batches corresponded to adults produced in hatchery from 2011 to 2013, grown in sea pens in the restocking Tiabet Bay and whose breeders came from a single site (*i.e.* site 1 or 4; batches A-C, Table 1). Samples were directly collected in the sea pens while being fished.
- One batch was composed of adults that accidentally escaped from damaged sea pens (batch D; Table 1). Batch D was an undetermined subsample of batches A-C (Table 1). This sample was also directly collected in the sea pen while being fished.

Batches A – D therefore represented the bay's new pool of breeders that survived from 2011 to 2013 and were pooled under the BAY label in subsequent analyses.

- The last batch corresponded to hatchery-produced juveniles whose breeders came from site 4 and that were sampled before their release in the restocking Tiabet Bay (batch E, Table 1).

A total of 1626 sandfish (400 wild adults and 1226 hatchery-produced individuals) has been sampled for subsequent genetic analyses.

## 2.2 DNA extraction and microsatellite genotyping

Genomic DNA was extracted using a cetyltrimethyl ammonium bromide (CTAB) protocol (Hillis and Moritz, 1990). Forward primers of *H. scabra* specific microsatellite loci (Fitch et al., 2013) were labelled with ABI fluorescent dyes and 16 loci were amplified into two multiplex PCRs as follows: Multiplex1: Hsc1/6-FAM, Hsc20/6-FAM, Hsc12/NED, Hsc42/NED, Hsc11/VIC, Hsc28/VIC, Hsc44/PET and Hsc62/PET; Multiplex2: Hsc4/6-FAM, Hsc49/6-FAM, Hsc17/NED, Hsc24/NED, Hsc48/VIC, Hsc59/VIC, Hsc31/PET and Hsc40/PET.

Multiplex PCR reactions were completed in a thermocycler (GeneAmp PCR System 2700, Applied Biosystems, Foster City, CA, USA) using the Type-it Microsatellite PCR Kit (Qiagen) in a final volume of 5 µL containing 2X Master Mix, 0.5X of Q-solution, 0.1µM of each primer (fluorescent-labeled forward primer 6-FAM, PET, NED or VIC) and 50 to 150ng of DNA template. Thirty-three PCR cycles were used with an annealing temperature of 57°C. Amplified fragments were sent to GENTYANE platform (INRA, Clermont-Ferrand, France), where they were resolved on an ABI 3730XL sequencer along with a GeneScan LIZ-500 internal size standard (Applied Biosystems). Alleles were sized using GENEMAPPER v. 4.0 (Applied Biosystems). We used GMCONVERT (Faircloth, 2006) to convert the exported GENEMAPPER table of genotypes.

## 2.3 Data Analysis

2.3.1 Genetic diversity analyses

Genetic diversity was estimated for each sample through (1) allele frequencies, (2) the average number of alleles ($N_{all}$), and (3) the average allelic richness per locus per sample ($A_r$) using the *DiveRsity* package (Keenan et al., 2013; R Core Team 2017). A second approach, implemented in the *ARES* package (van Loon et al., 2007), was also used to estimate allelic richness: while the first approach estimates the expected number of alleles corrected for sampling size based on a rarefaction procedure, the second approach extrapolates (and/or rarifies) the allelic richness beyond the sampling size and provides confidence intervals for any sample size. A sample size up to 100 individuals was used to estimate the allelic richness using the *ARES* package ($A_{r-ARES}$). Genetic diversity was also assessed through expected heterozygosity ($H_e$) and fixation index ($F_{IS}$) using GENEPOP 4.1 (Rousset, 2008), which was also used to test for departures from Hardy–Weinberg equilibrium in each sample (10 000 dememorization steps, 500 batches and 5 000 iterations per batch).

2.3.2 Genetic population structure analyses

The genetic structure among samples was assessed by calculating an estimate of $F_{ST}$ (Weir and Cockerham, 1984) using GENEPOP 4.1. Exact tests for population differentiation (10 000 dememorization steps, 500 batches and 5 000 iterations per batch) were carried out to test for differences in allele distributions across samples and between pairwise samples. Then, we estimated the assignment of individuals to the different genetic clusters using the individual-based Bayesian clustering method implemented in STRUCTURE v. 2.3.4 (Falush et al., 2003; Pritchard et al., 2000). For each value of the K parameter (ranging from 1 to 7), five replicate chains of 500 000 Markov Chain Monte Carlo (MCMC) iterations were run after discarding 50 000 burn-in iterations. STRUCTURE was first run using the natural populations (sites 1 - 4, Fig. 1) using default parameters. As that approach leads to erroneous results when populations are unevenly sampled (Puechmaille, 2016; Wang, 2017), an admixture model with uncorrelated allele frequencies and the alternative prior was applied when adding hatchery-produced samples. Following Wang (2017)'s recommendations, several models were tested, in which the initial alpha parameter varied from 0.15 (c.a. ⅙, if K=6) to one (K=1, suggesting one panmictic population). Since varying alpha parameter led to very similar results, results with alpha = 0.5 (½, with a "true" K=2) were thereafter presented. To determine individual ancestry proportions that best matched across all replicate runs, CLUMPP (Jakobsson and Rosenberg, 2007) was used and individuals' assignment was visualized using the R software.

The use of multiple methods for determining genetic structure is usually advocated, STRUCTURE making strong priors and hypotheses that are usually challenged (e.g. requirements of Hardy-Weinberg and Linkage equilibria; Jombart et al., 2010). In addition to $F_{ST}$ estimates and in order to graphically visualize the overall genetic distance between samples, the genetic structure among sampling sites was also depicted using Principal Component Analysis (PCA) was performed using the R package ADEGENET 1.4–2 (Jombart, 2008; Jombart et al., 2011), a distance-based approach for which few (nearly no) assumptions may be violated. In addition, the model-free Discriminant Analysis of Principal Components, DAPC (Jombart et al., 2010), which is also a multivariate genetic clustering method that does not make use of linkage equilibrium or Hardy-Weinberg Equilibrium, or any genetic model, was applied as well.

2.3.3 Kinship analyses

To investigate whether released hatchery-produced individuals are more related one to each other than with wild individuals, relatedness coefficients between all pairs of individuals were computed using the approach described in Wang (2007, 2011). Briefly, this approach uses maximum likelihood to infer the relatedness coefficient between two individuals given population allele frequencies.
As a prerequisite, we tested the performance of the six different estimators described by Wang (2011), aiming to choose with accuracy the relatedness estimator that most suited our dataset. We simulated individuals of known relatedness (e.g. fullsibs or parent–offspring should have a relatedness coefficient of 0.5 in a large random-mating population) using the same locus characteristics than our dataset. Based on these simulations, none of the six estimators of relatedness proposed by Wang (2011) consistently performed better than another (Fig. SI1). Relatedness coefficients were computed with the triadic likehood estimator (triML) estimator in the subsequent analyses, which showed the highest correlation coefficient among the six tested estimators (Supporting Information 1). Simulations based on all genotypes were also performed prior to the estimation of relatedness coefficients based on the real dataset to assess variation in relatedness coefficients for four categories targeted in the analysis (full-sibs, half-sibs, parent-offspring and unrelated individuals). The distributions of simulated relatedness coefficients overlapped between the four targeted categories, making it difficult to unambiguously assign a pair of individuals as full-sibs, half-sibs, parent-offspring or unrelated individuals (SI2). Simulated unrelated, half-sibs, full-sibs, and parent-offspring individuals were defined by a relatedness coefficients ranging from 0 to 0.55, 0 to 0.89, 0.02 to 1, and 0.06 to 1, respectively. Pairs with a relatedness coefficient higher than 0.55 were then unambiguously identified as related individuals.
Finally, relatedness estimates, mean relatedness coefficients and their 95% confidence intervals were estimated based on our dataset with a bootstrap procedure (500 iterations). All these analyses were performed using the *related* package (Pew et al., 2015).

2.3.4 Inbreeding analyses

To further test for possible inbreeding occurring within samples of hatchery-raised juveniles, individual Multi-Locus Heterozygosity (MLH) was calculated for each individual. MLH is defined as the number of heterozygote locus standardized by the number of loci genotyped (David et al., 2007; Szulkin et al., 2010). In a population with variance in inbreeding, inbred individuals are less heterozygous (*i.e.* lower MLH). Inbreeding variance generates Identity Disequilibria (ID) -*i.e.* correlations in homozygosity across loci, a measure of departure from random associations between loci (Szulkin et al., 2010). IDs were measured for each sampling site by calculating the $g_2$ parameter and its standard deviation using the RMES software (David et al., 2007). The $g_2$ parameter measures the excess of double heterozygotes at two loci relative to the expectation under a random association, standardized by average heterozygosity (Szulkin et al., 2010), providing a measure of genetic association and inbreeding variance in the population. To test for the significance of $g_2$, random re-assortments of single-locus heterozygosities among individuals were tested using 1 000 iterations (David et al., 2007).

2.3.5 Effective number of breeders ($N_b$) estimation

The effective number of breeders that produced the sample $N_b$ (Waples, 2005) was estimated within each sample. We measured the non-random association of alleles at different loci within F1 population (Hill, 1981) using a bias correction (Waples, 2006; Waples and Do, 2008) implemented in NeEstimator v2 software (Do et al., 2014). Following Waples and Do (2010)'s recommendations, we excluded alleles with frequencies below 0.02 and random mating was applied according to *H. scabra* reproductive characteristics.

**3. Results**

**3.1 Heterogeneous genetic diversity among sampling sites**

From the originally amplified 16 loci, seven loci were removed from downstream analyses due to either high frequencies of missing genotypes (indicative of null homozygotes caused by null alleles) or multiple peaks profile. A total of 1358 individuals was successfully genotyped at nine polymorphic loci (9.28 % of missing data, two to 32 alleles/locus, Table 2). Mean allelic richness per locus per sample ($A_r$) and expected heterozygosity ($H_e$) were homogeneous across samples, ranging from 4.04 (batch B) to 6.59 (site 4) and from 0.453 (batch B) to 0.581 (site 4), respectively (Table 2). Allelic richness ($A_{r-ARES}$) varied from 53.7 (95% CI = [42.4 - 67.0], batch B) to 78.5 (95% CI = [62.2 - 86.8], site 4, Table 2) at an extrapolated/rarefied sample size of 100. Allelic accumulation curves revealed no significant difference across wild samples (*t*-test *p*-values > 0.05, Fig. 2a). However, when comparing the genetic diversity of hatchery-produced samples to that of their respective breeder population(s), we found that the total number of distinct alleles significantly differed between site 1 and batch B (*t*-test *p*-values < 0.05 at a minimum sample size of 24, Fig. 2b and 2d) and between site 4 and batches C and E, and BAY (*t*-test *p*-values < 0.05 at a minimum sample size of 22, 1 and 53, respectively, Fig. 2c and 2e) as the extrapolated/rarefied sample size increased.

**3.2 Weak but significant genetic structure between wild and released *H. scabra***

Distance-based analyzes (PCA and DAPC) did not reveal any genetic differentiation across all sampling sites, which was consistent with the observed genetic panmixia among all sampling sites ($F_{IS-wild\ populations}$ = 0.080 ns, $F_{IS-all\ populations}$ = 0.091 ns). Only site 2 slightly, though significantly, differed from the three other wild populations ($F_{ST}$ = 0.003 - 0.006, Table 3). No wild individual was completely assigned to any new cluster based on STRUCTURE, but rather to an equal admixture for each individual within a cluster (data not shown). However, running STRUCTURE with all populations (*i.e.* wild and released populations) revealed a different genetic pattern. The K value of 3 provided the best meaningful result with most individuals of batch E assigned to an additional cluster (Fig. 3). Further increases in K did not result in meaningful geographic clusters. $F_{ST}$ analyses converged toward very similar results, indicating weak, yet significant genetic differentiation between batch E and all other sites (from 1% to 2%), and to a lesser extent, between hatchery-produced samples and all other sites (Table 3).

**3.3 Slightly higher relatedness in released populations than in wild populations**

In each sample, relatedness estimates ranged from 0, *i.e.* unrelated individuals, to 1, *i.e.* unambiguously highly related individuals. Most relatedness estimates were however lower than 0.125, meaning that most individuals were unrelated (from 63.61 % in batch A to 74.96 % in site 4, Fig. 4a, without taking batch B into account (see below)). For the relatedness estimates ranging from 0.125 to 0.25, *i.e.* considered as halfsibs, avuncular or a grandparent-grandchild relatedness, from 13.36 % (site 4) to 16.47 % (site 2) were observed (Fig. 4b). Finally, less than 2.5 % (0.46 % in site 2 to 2.30 % in batch D, Fig. 4d) were identified with a relatedness coefficient higher than 0.55, *i.e.* the simulated value for which pairs were unambiguously identified as related. Interestingly, batch B showed a different pattern: 33.04 % of relatedness estimates were identified to unrelated dyads while 20.87 % of dyads corresponded to a relatedness coefficient higher than 0.55 (*i.e.* unambiguously identified as related individuals). Overall, the mean relatedness coefficients ranged from 0.087 (site 2) to 0.317 (batch B) and were not significantly different between wild populations and hatchery-produced batches (bootstrap *p*-value > 0.5), with the exception of batch B.

**3.5 Inbreeding and small effective number of breeders ($N_b$) in released populations**

Individual Multi – Locus Heterozygosity (MLH), *i.e.* the number of heterozygote locus standardized by the number of genotyped loci, ranged from 1 (batches D and E) to 8 (all sampling sites, except site 2, and batches B and C). MLHs were normally distributed (Shapiro test *p*-values > 0.5) and no difference was observed in MLH distributions (ANOVA *p*-value > 0.5). Identity disequilibrium, *i.e.* a measure of departure from random associations of homozygosity between loci, was identified in four samples, in site 2, in batch D, in batch E, as well as at the BAY scale (Table 2), suggesting that few inbred individuals were present in those samples. All hatchery-produced samples showed a clear reduction in the effective number of breeders ($N_b$ ≥ 90%), which greatly contributed to the loss of genetic variation among sandfish juveniles (Tables 2 and 4). While large point estimates ($N_b$ > 300, with infinite CIs) were inferred in wild populations (Table 2), smaller values of $N_b$ ($N_b$ < 55) were estimated in hatchery-produced samples, especially in batch B ($N_b$ = 6.9, CI = [3.6, 11.5]).

**3.6 Cross-analysis of genetic indicators within the hatchery-produced samples**

When comparing population genetic estimates among the five batches of hatchery-produced individuals, batch B displayed the smallest $N_b$ point estimate, the strongest decrease in genetic diversity, the strongest increase in highly related individuals, and the most important structure when compared to the breeders' site (Table 4). Interestingly, batch A showed the less reduction in $N_b$ point estimate, exhibited the less difference in genetic diversity, as well as the less highly related dyads when compared to the breeding site, while it displayed a similar sampled size than batch B (Table 4). Further, despite the largest sampled size in batch E, all genetic indicators were lower in that batch than in the breeders' population, in the same order of magnitude than batch B. Batches of hatchery-produced individual with several breeders' origin did not change the decrease in all genetic indicators (e.g. batch D and BAY; Table 4).

**4. Discussion**

**4.1 Setting the scene for restocking events: genetic diversity and structure of wild sandfish stocks**

Using nine microsatellite markers, we found that wild sandfish individuals were genetically homogeneous from site 1 to site 4, *i.e*. over ca. 90 km. This result suggested a large and panmictic population over the northwestern coastline of New Caledonia, as previously observed in the study species by Uthicke and Purcell's (2004) using seven allozyme markers. Restocking, stock enhancement, and fishing regulations of sandfish may therefore be appropriately defined at that scale. The genetic homogeneity observed based on nine microsatellite markers is consistent with the estimated genetic connectivity patterns of *H. scabra* populations at similar scales (ca. 100km) outside New Caledonia: in Northern Australia (Gardner et al., 2012), Papua New Guinea (Nowland et al., 2017), Philippines (Ravago-Gotanco and Kim, 2019), and Thailand (Ninwichian and Klinbunga, 2020). Although the oceanographic features of the studied areas may differ and should be carefully taken into account, all these studies were however pointing towards genetic homogeneity over a 100km scale. This may question, at the first site, the microsatellite power resolution. Indeed, the four above-mentioned studies used the same microsatellite markers specifically developed for *H. scabra* (Fitch et al., 2013) as those of the present study. The set of markers used were however able to detect significant genetic structure among wild populations, although at larger spatial scales. For instance, significant genetic differentiation was observed along the Adaman coast of Thailand, at a 240-360 km study scale (Ninwichian and Klinbunga, 2020), suggesting that this set of markers seems thus likely reliable to observe genetic structure. Moreover, despite the genetic homogeneity observed from site 1 to site 4 in the present study, evidence of weak, although significant, population structure among wild populations was found at small spatial scale here (*i.e.* site 2), possibly due to the geomorphology of the semi-enclosed bay where samples were collected. On the one hand, accordingly to the relationship between dispersal capacities and pelagic larval duration of marine species (Shanks et al., 2003), the pelagic larval duration of 10-14 days of that species (Hamel et al., 2001) seemed likely consistent with the genetic homogeneity observed from site 1 to site 4. On the other hand, the existence of hydrological and ecological barriers, such as currents, temperature, or salinity (Durand et al., 2004) as well as species' life history traits (e.g. larval behavior; Whitaker, 2004) may limit gene flow, sometimes over relatively short distances (e.g. Shanks, 2009). Local retention of water masses in site 2 may limit water exchanges with surrounding habitats and therefore impede larvae movements and consecutive gene flow among local sandfish populations, despite a pelagic larval duration of 10-14 days .

The effective number of breeders ($N_b$) in the wild sandfish populations in sites 1 and 4 was estimated with infinite upper confidence limit, large, and of the same order of magnitude as the effective population size estimated in the Philippines archipelago (Ravago-Gotanco and Kim, 2019). Although $N_b$ estimation accuracy should be cautiously interpreted (Hare et al., 2011), such large $N_b$ estimates would likely ensure that genetic variation is maintained within these populations in long term, provided that sustainable management practices are enforced. Specifically, following Frankham et al. (2014), $N_b$ estimates were close or higher than the reference threshold ($N_b$= 1 000) that would allow for maintaining evolutionary potential in perpetuity in both sites (sites 1 and 4), while lower limit of $N_b$ confidence intervals, *i.e.* an indicator of the lowest possible level of $N_b$ (Hare et al., 2011), was higher than the critical threshold of inbreeding depression ($N_b$= 100) in sites 2 and 3. Because panmixia was observed among wild populations from site 1 to site 4 -site 2 significantly differing from these sites-, $N_b$ estimates would be closer to a metapopulation size of contemporary generations than the local effective population size ($N_e$, Palstra and Ruzzante, 2008; Waples, 2005), which would partly explain the high effective population size estimates in the study area. Although we used a unique approach to estimate $N_e$ while estimations would gain in confidence by using several approaches, the high $N_b$ estimates appeared not supportive of the worldwide IUCN endangered status of the sandfish. Such a high effective number of breeders would theoretically allow for developing hatchery procedures and subsequent restocking and stock enhancement programs that can maintain wild genetic diversity if properly managed (see section 4.2).

Effective population size ($N_e$) should however be interpreted in comparison to census population size ($N_c$) in the conservation management context. The $N_e/N_c$ ratio is indeed important for disentangling the relative risks that demographic, environmental and genetic factors might pose for population persistence in short term (Frankham, 1995; Palstra and Ruzzante, 2008). The theoretical $N_e/N_c$ ratio is expected to be ca. 0.5 for a wide range of species displaying different demographic and reproductive life traits (Nunney, 1993, 1996), a ratio reevaluated to 0.1 (Frankham, 1995). However, this ratio may greatly decrease for species displaying high fecundity and high mortality in the early stages (Palstra and Ruzzante, 2008), such as encountered in many marine invertebrates (Eldon et al., 2016; Hedgecock, 1994), and sea cucumbers in particular. In case of a very low ratio, *i.e.* when only a few breeders effectively reproduce, the $N_e/N_c$ ratio may be approximated by $N_b/N_c$ (Hedrick, 2005). In this study, the census population sizes ($N_c$) were estimated at around 500 000 and 220 000 individuals in sites 3 and 4, respectively, based on previous surveys by Léopold et al. (2013, 2015). $N_c$ is thus three orders of magnitude higher than the effective population size in both cases ($N_b/N_c$ =0.6 $10^{-3}$ in site 3 and 4.4 $10^{-3}$ in site 4). Such very low, estimated $N_e/N_c$ ratios were therefore typical of species exhibiting high variance in reproductive success (Eldon et al., 2016; Hedgecock et al., 1994), but also of species displaying fluctuations in population size through time (Nunney, 1993; Vucetich et al., 1997). It should then be carefully taken into account in sandfish population restocking programs as an indicative parameter for tracking temporal trends in wild populations and that rapid and low-coast genetic monitoring can be used to detect fluctuations in abundance.

**4.2 Lowering the genetic risk of diversity alteration in hatchery-produced sandfish population**

In the present study, five interrelated genetic indicators ($A_r$, $F_{ST}$, relatedness coefficient, inbreeding coefficient, and effective number of breeders) provided relevant insights for restocking and stock enhancement of *H. scabra*. During any of the spawning event in hatchery for the present study, estimated $N_b$ was lower than the critical threshold defined by Frankham et al. (2014), which increases the likelihood of extinction of hatchery-produced populations in the short term. This finding means that the genetic diversity of hatchery-produced sandfish, lowered by ca. 5-35% when compared to wild populations, was too small to prevent evolutionary potential loss and, ultimately, population extinction if these individuals were grown-out for restocking or stock enhancement. This is likely attributed to the typically-small number of breeders that effectively reproduce in fish and invertebrate hatcheries (Araki et al., 2007; Boudry et al., 2002; Chapman et al., 2002; Lemay and Boulding, 2009; Rourke et al., 2009), and particularly in sandfish. Indeed, Gardner et al. (2012) showed that only five out of the 40 breeders were effectively reproducing, being responsible for around 71% of the descendants. The present genetic study is the first to assess the risk of genetic diversity alteration in sandfish restocking and stock enhancement programs and highlighted some weaknesses associated with hatchery-produced samples for restocking, *i.e.* the important decrease in genetic diversity that will detrimentally affect the released populations. To maximize the effective number of breeders ($N_b$) and subsequent genetic variability of hatchery-produced juveniles, particular attention must therefore be paid to broodstock management, not only at the initial stages of sandfish restocking and stock enhancement programs, but also at each step during and after reseeding. Large and random subsets of breeders should be collected within the wild panmictic sandfish population to capture a high proportion of the natural genetic diversity as proposed by Purcell et al., (2012), Quinn (2005) and Waples (1991), among others. Moreover, the exchange of broodstocks between hatcheries, the distribution of artificial seedlings to different regions, as well as genetic rescue in the field, can be additional guidance in hatchery-produced samples for restocking. Finally, given the high variance in reproductive success of spawning inducement events and the uncontrolled natural selection during spawning and at early-life stages, long term and appropriate genetic monitoring should be conducted to ensure that genetic variation is maintained throughout the restocking steps and after reseeding.

## 5. Conclusion

Using nine microsatellite markers, our study of *H. scabra* population genetics in New Caledonia revealed one large and panmictic stock in the northwest lagoon area. Restocking and stock enhancement programs and fishery regulations may therefore be defined at that spatial scale. Small-scale genetic variation was however detected in a restricted area, suggesting that locally-based resource management may be suitable in specific environmental context. The relative low resolution of microsatellite markers, thought proven to be efficient for identifying in the present study spatial scales of genetic structure in site 2, would prevent to identify subtle pattern of very low genetic differentiation, difficult to grasp with the limited statistical power of our dataset. To get higher resolution, the easiest way would be to increase the number of markers, e.g. the use of RAD-associated SNPs, which also would allow gaining in confidence about $N_e$ estimates (Luikart et al. 2010) and investigating the genetic load (Plough 2016) in a conservation frame. Genetic diversity loss was however found within hatchery-produced populations, which would predictably lead to detrimental effects of genetic drift and inbreeding. Hatcheries should therefore define wild broodstock collection and management procedures to maintain genetic diversity of hatchery-produced stock. Further genetic implications of restocking and stock enhancement programs should be monitored to determine the impacts of hatchery-produced populations on neighbor wild populations.

**Acknowledgments**

We are grateful to Claire Marty, Zacharie Moenteapo, Loïc Bourgine, Yoané Tein-Bai, Jérôme Azzaro, and Jean-François Kayara from the Fisheries Division of the Northern Province of New Caledonia for their support for fieldwork, maintenance of enclosures for sandfish, and facilitating relationships with the Tiabet community. We thank the Tiabet's Kanak chief, fishers and other community members for their kindness and commitment during the 3-year project. We also thank Gérard Mou-Tham, Christophe Peignon and Joseph Baly from the IRD center of Noumea for their technical contribution, as well as the SEA (Société d'Elevage Aquacole) sandfish hatchery at Bouloupari, New Caledonia for their collaboration. This study was supported by the Northern Province of New Caledonia as part of the RESCA N°14C212 grant and by the Laboratoire d'Excellence CORAIL (French National Research Agency) through a postdoctoral fellowship.

**Author contributions**

**Florentine Riquet:** Data analyses, Writing- Original draft preparation; **Cécile Fauvelot:**

Conceptualization, Writing- Reviewing and Editing; **Pauline Fey:** Sampling and molecular analysis;

**Daphné Grulois:** Genotyping**,** Molecular analysis; **Marc Léopold:** Conceptualization, Writing -

Reviewing and Editing. All authors read and improved the manuscript.

**Data availability**

Microsatellite data has been deposited at DRYAD: *to be completed after the manuscript is accepted for publication*

**Tables**

**Table 1.** Batches characteristics of hatchery-produced sandfish juvenile that were sampled for further genetic analyses.

| Batch ID | N collected | Release area | Release date | Breeder's site |
|---|---|---|---|---|
| A | 50 | Sea pen (700 m²) | October 2011<br>July 2013 | site 1 |
| B | 50 | Sea pen (1700 m²) | December 2013 | site 1 |
| C | 50 | Sea pen (2900 m²) | October 2012 | site 4 |
| D* | 76 | Sea pen | October 2011<br>to December 2013 | sites 1 and 4 |
| E | 1000 | Sea pens (13800 m²) then open habitats | June to October 2014 | site 4 |

* Batch D was composed of an unknown percentage of juveniles of sea pens A, B, and C that escaped following sea pen damage and were captured in neighbor open habitat.

**Table 2.** Genetic diversity indices of the study samples using nine microsatellite markers. N: number of individuals, $A_r$: allelic richness, $A_{r100}$: allelic richness rarefied/extrapolated at a sampling size of 100 and its 95% CI, $H_e$: expected heterozygosity, $F_{IS}$: fixation index (p-values of the exact test associated with Hardy-Weinberg non-significant for all samples except for site 2 where a deficit in heterozygotes is observed, $F_{IS} > 0$), $g_2$: estimate of the identity disequilibrium. Probability values for $g_2 = 0$ were corrected for multiple comparisons: (non-significant, p-value >0.05), * (p-value <0.05), ** (p-value <0.01), *** (p-value <0.001). $N_b$: effective number of breeders and its 95% CI. A total of 40 individuals were used to estimate $A_r$, which took into account variation of sample size, when examining $A_r$ of all samples, and 80 individuals were used when examining $A_r$ of the BAY and all other samples ($A_r$ given in brackets). Note that $A_r$ of the Bay is only estimated in the letter case, while $A_r$ of A-, B-, C-, D- and E-batches in the former case.

| site | N | $A_r$ | $A_{r100}$ | $CI_{r100}$ | $H_e$ | $F_{IS}$ | $g_2$ | $N_b$ | $CI_{-Nb}$ |
|---|---|---|---|---|---|---|---|---|---|
| *Wild populations* | | | | | | | | | |
| 1 | 98 | 5.66 (6.61) | 69.4 | [62.3 - 79.0] | 0.536 | 0.074 | 0.008 | ∞ | [914.3, ∞[ |
| 2 | 94 | 6.46 (7.49) | 77.1 | [70.3 - 90.3] | 0.572 | 0.185* | 0.035* | 369.4 | [132.9, ∞[ |

| | | | | | | | | |
|---|---|---|---|---|---|---|---|---|
| 3 | 91 | 6.46 (7.54) | 77.6 | [74.2-84.7] | 0.572 | 0.063 | 0.02 | 312.1 | [117.3, ∞[ |
| 4 | 97 | 6.59 (7.65) | 78.5 | [62.2 - 86.8] | 0.581 | -0.001 | -0.014 | 969.7 | [192.9, ∞[ |
| *Hatchery produced juveniles* | | | | | | | | | |
| **BAY** | 207 | (6.76) | 64.6 | [56.6 - 72.7] | 0.550 | 0.007 | 0.016* | 52.2 | [41.9, 65.5] |
| **A** | 49 | 5.36 | 65.2 | [54.4 - 77.8] | 0.544 | -0.053 | 0.020 | 91.9 | [41.5, 2699.1] |
| **B** | 46 | 4.04 | 53.7 | [42.4 - 67.0] | 0.453 | -0.112 | -0.010 | 6.9 | [3.6, 11.5] |
| **C** | 43 | 4.91 | 54.1 | [47.0 - 66.4] | 0.569 | -0.013 | 0.018 | 38.2 | [21.4, 95.3] |
| **D** | 69 | 5.54 | 71.7 | [60.8 - 86.9] | 0.546 | 0.016 | 0.027* | 62.4 | [38.0, 128.8] |
| **E** | 771 | 5.83 (6.56) | 58.5 | [50.4 - 66.7] | 0.568 | 0.105 | 0.024 *** | 31.1 | [27.4, 35.1] |

**Table 3**. Genetic structure (pairwise F$_{ST}$ estimates and probability values from exact tests) between samples. *P*-values were all significant (*p*-value <0.001, after Bonferroni correction), except for F$_{ST}$ estimates in bold.

|     | 1 | 2 | 3 | 4 | BAY | A | B | C | D |
|---|---|---|---|---|---|---|---|---|---|
| 2   | 0.004 |       |       |       |       |       |       |       |       |
| 3   | **0.000** | 0.006 |       |       |       |       |       |       |       |
| 4   | **0.000** | 0.003 | **0.000** |       |       |       |       |       |       |
| BAY | 0.005 | 0.013 | 0.005 | 0.006 |       |       |       |       |       |
| A   | 0.007 | 0.017 | 0.007 | 0.009 |       |       |       |       |       |
| B   | 0.077 | 0.082 | 0.079 | 0.075 |       | 0.085 |       |       |       |
| C   | 0.014 | 0.016 | 0.011 | 0.008 |       | 0.020 | 0.121 |       |       |
| D   | 0.005 | 0.012 | 0.002 | 0.007 |       | **0.003** | 0.073 | 0.018 |       |
| E   | 0.012 | 0.014 | 0.017 | 0.014 | 0.019 | 0.016 | 0.083 | 0.023 | 0.021 |

**Table 4**. Comparison of the genetic diversity, structure and kinship of restocking events regarding the breeding site

|     | N | Breeding site | Difference in N$_b$ point estimates with the breeding site* (%) | Difference in A$_{r-100}$ with the breeding site (%) | F$_{ST}$ with the breeding site** | Difference in highly related dyads (r>0.5, %) |
|---|---|---|---|---|---|---|
| A   | 49  | 1 | -89.95 | -6.12  | 0.007 | -10.71 |
| B   | 46  | 1 | -99.25 | -36.26 | 0.077 | +1053.41 |
| C   | 43  | 4 | -96.06 | -31.14 | 0.008 | +68.36 |
| E   | 771 | 4 | -96.79 | -25.48 | 0.014 | -4.77 |
| BAY | 207 | 1 | -94.29 | -6.88  | 0.005 | +33.24 |
|     |     | 4 | -94.62 | -17.71 | 0.006 | +129.07 |
| D   | 69  | 1 | -93.18 | +3.33  | 0.005 | +27.21 |
|     |     | 4 | -93.57 | -8.70  | 0.007 | +118.72 |

*As N$_b$ point estimate of site 1 is infinite, we used the lower bound of the confident interval when comparisons involved this point estimate

** *p*-values < 10$^{-3}$ for all comparisons

**Figures caption**

**Figure 1:** Sampling locations of *Holothuria scabra* along the western coast of the North Province of New Caledonia (a), with location of the study area in New Caledonia (b). Each study site is labeled as follow: 1- Koumac, 2- Tinip, 3- Boyen, 4- Kone for wild populations and A, B, C, D and E for hatchery-produced samples released in Tiabet, represented by a black star on the map. A-, B-, C- and D-batches representing the bay's new pool of breeders were pooled under the BAY label.

**Figure 2:** Allelic accumulation curves of *Holothuria scabra* of wild populations (a), population at site 1 and all samples implying breeders originated from site 1 (b), and population at site 4 and all samples implying breeders originated from site 4 (c). *P*-values when testing the differences in mean allelic richness are presented between site 1 and all samples implying breeders originated from site 1 (d) and between site 4 and all samples implying breeders originated from site 4 (e). Curves are colored following the hatchery-produced samples (Fig. 1). When a curve falls below the 0.05 horizontal line (grey dotted line), the null hypothesis of equal richness is rejected with a 0.05 confidence level.

**Figure 3:** Individual Bayesian ancestry proportions determined using STRUCTURE for wild populations and released samples (A-, B-, C-, D-, E-batches) with K=3 clusters identified. Dotted black lines separate each study site. Sites are labeled as in Fig. 1. Each individual is depicted as a vertical bar with colors distinguishing its ancestries to the three clusters.

**Figure 4:** Frequency of pairs of individuals distribution with a relatedness ranging from 0 to 0.125 (a), 0.125 to 0.25 (b), 0.25 to 0.5 (c) and above 0.5 (d). Each study site is colored and labeled following Fig. 1.

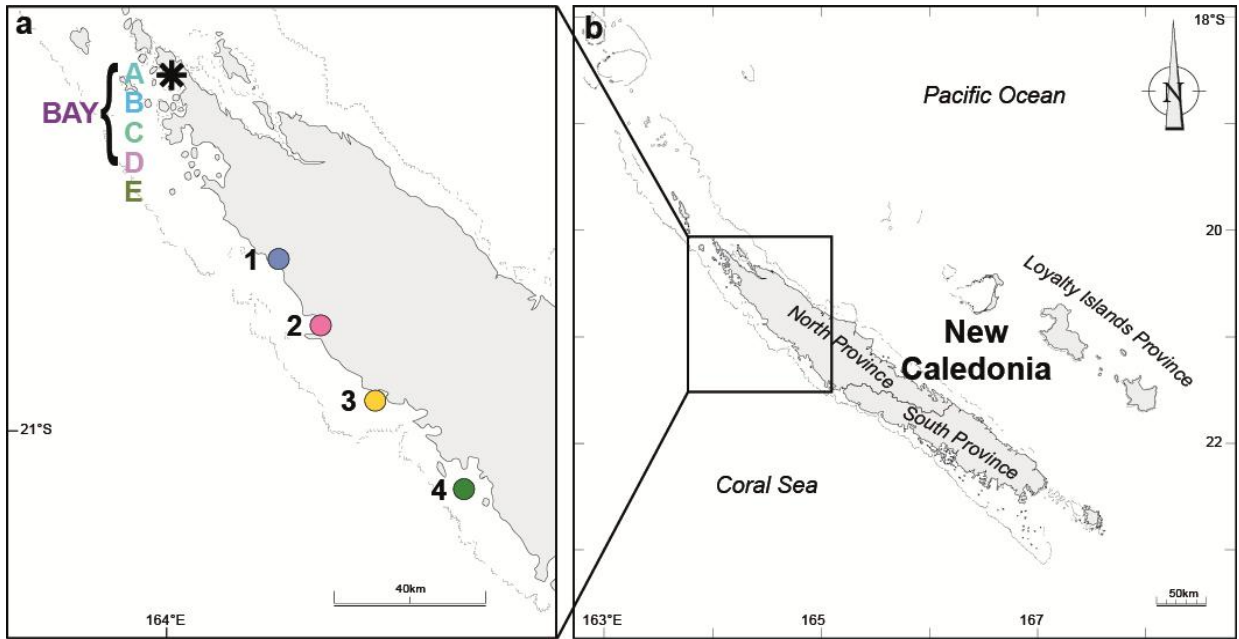

**Figure 1**

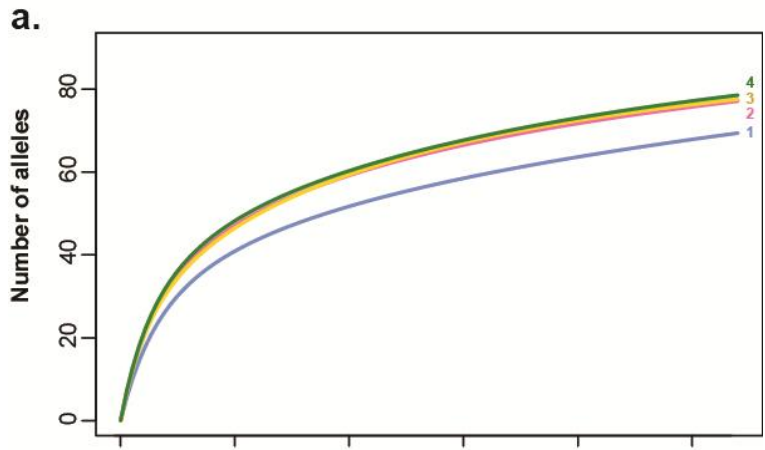
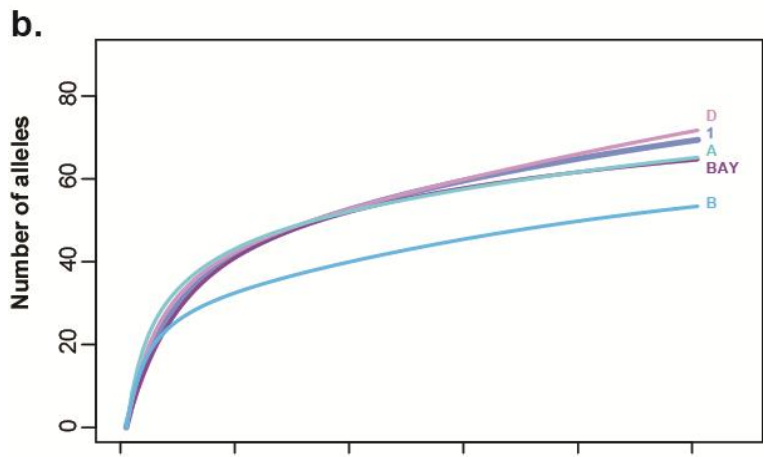
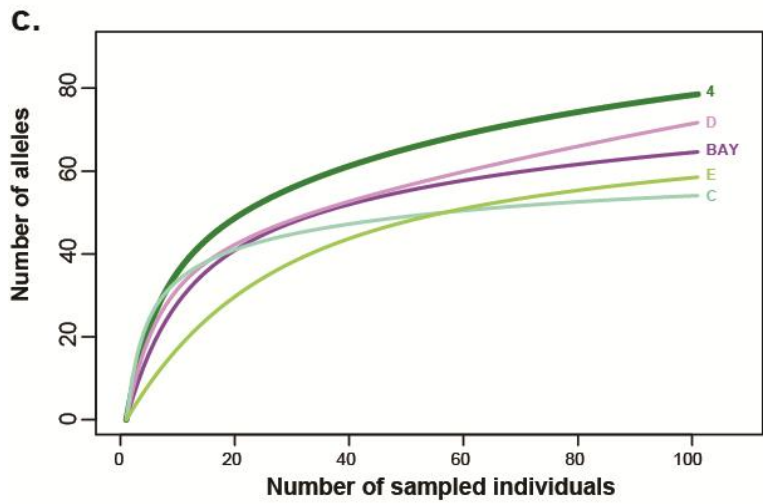
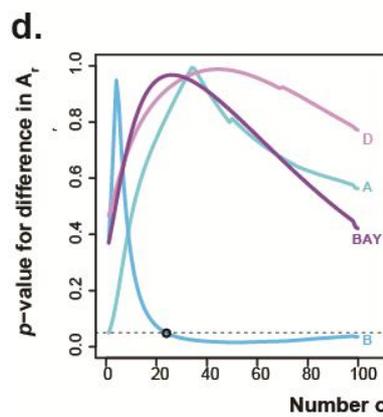
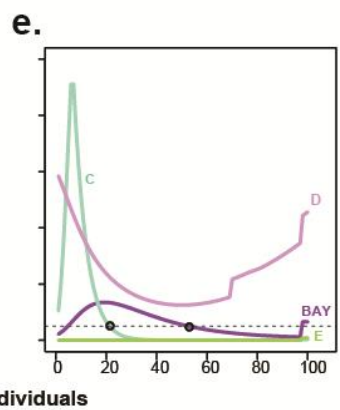

**Figure 2**

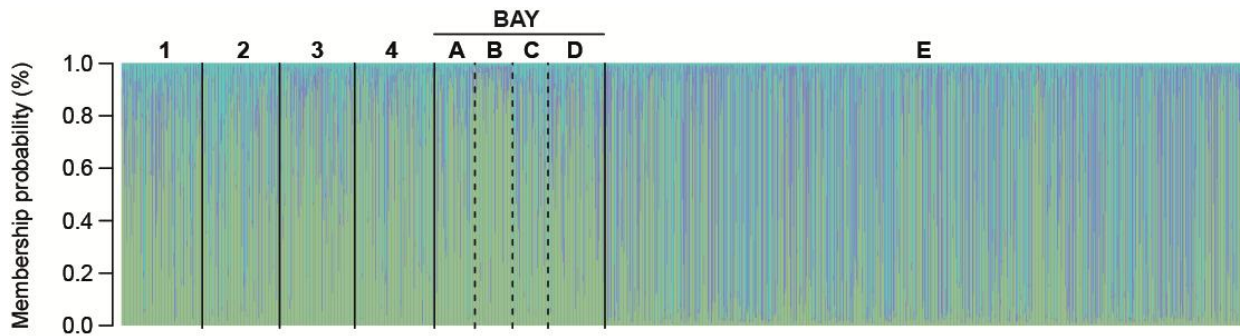

**Figure 3**

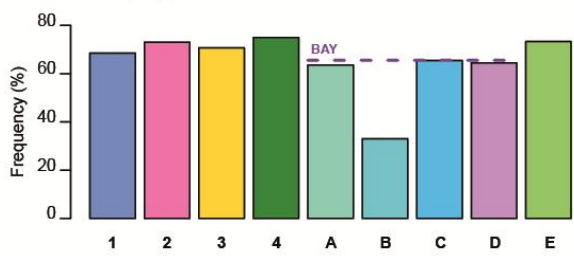
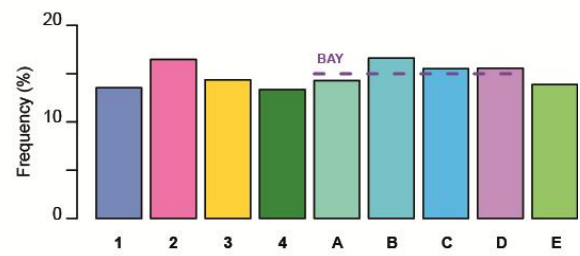
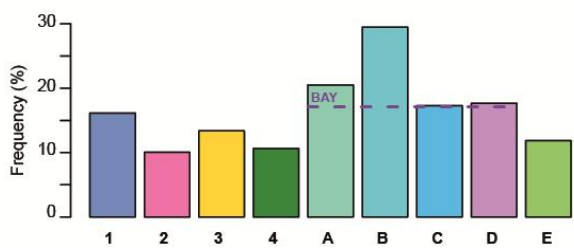
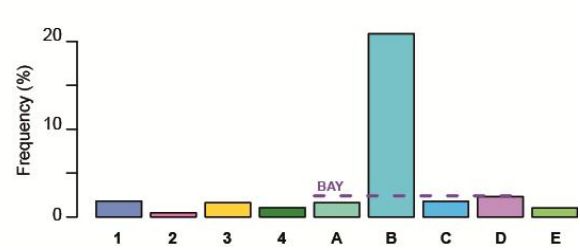

**Figure 4**